\documentclass[conference]{IEEEtran}
\IEEEoverridecommandlockouts
\usepackage{epsfig}
\usepackage[latin1]{inputenc}
\usepackage[english]{babel}
\usepackage{graphicx}
\usepackage{psfrag}
\usepackage{amsmath,amsfonts,amsthm,amsxtra,amssymb,mathrsfs}   
\usepackage{lscape}
\usepackage{bm}
\usepackage{verbatim}
\usepackage{bbm}

\title{Spectrum Sensing in Cognitive Radios Based on Multiple Cyclic Frequencies\\
{\em \normalsize Invited Paper}}

\author{\IEEEauthorblockN{Jarmo Lund\'en\thanks{Jarmo Lundén's work was supported by GETA graduate school, Finnish Defence Forces Technical Research Centre and Nokia Foundation.}\IEEEauthorrefmark{1},
Visa Koivunen\thanks{The funding for Visa Koivunen's 
sabbatical term at Princeton University was provided by the Academy of Finland.}\IEEEauthorrefmark{1}\IEEEauthorrefmark{2},
Anu Huttunen\thanks{Anu Huttunen in on a research leave from Nokia Research Center.}\IEEEauthorrefmark{1}
and H. Vincent Poor\thanks{H. Vincent Poor's work was supported by the US National Science Foundation under Grants ANI-03-38807 and CNS-06-25637}\IEEEauthorrefmark{2}}
\IEEEauthorblockA{\IEEEauthorrefmark{1} SMARAD CoE, Signal Processing Laboratory \\
Helsinki Univ. of Technology, Finland\\
Email: jrlunden@wooster.hut.fi\\}
\IEEEauthorblockA{\IEEEauthorrefmark{2}School of Engineering and 
Applied Science,\\ Princeton University\\
Email: poor@princeton.edu}}

\begin{document}
 
\maketitle

\begin{abstract}
Cognitive radios sense the radio spectrum in order to find unused frequency bands and use them in 
an agile manner. Transmission by the primary user must be detected reliably even in the low signal-to-noise ratio (SNR) regime and in the face of shadowing and fading. Communication signals are 
typically cyclostationary, and have many periodic statistical properties related to the symbol rate, the coding and modulation schemes as well as the guard periods, for example.
These properties can be exploited in designing a detector, and
for distinguishing between the primary and secondary users' signals.
In this paper, a generalized likelihood ratio test (GLRT) for detecting the presence of cyclostationarity using multiple cyclic frequencies is proposed. Distributed decision making is employed by combining the quantized local test statistics from many secondary users. User cooperation allows for mitigating the effects of
shadowing and provides a larger footprint for the cognitive radio system.
Simulation examples demonstrate the resulting performance gains in the low SNR regime and the 
benefits of cooperative detection. 
\end{abstract}
 
\section{Introduction}

Spectrum sensing is needed in cognitive radios in order to find opportunities for 
agile use of spectrum. Moreover, it is crucial for managing the level of interference caused 
to primary users (PUs) of the spectrum. Sensing provides awareness of the radio operating 
environment. A cognitive radio may then adapt 
its parameters such as carrier frequency, power and waveforms dynamically in 
order to provide the best available connection and to meet the user's needs 
within the constraints on interference.

In wireless communication systems we typically have some knowledge on the waveforms and structural or statistical properties of the signals that the primary user of the spectrum is using. 
Such knowledge may be related to the modulation scheme, 
the symbol or chip rate of the signal, the channel coding scheme, training or pilot signals, 
guard periods, and the power level or correlation properties of the signal, just to mention a few. 
These properties may be used to design a detector that works in a very low SNR regime 
and has low complexity and consequently low power consumption. These are very desirable properties especially for cognitive radios in mobile applications. In the absence of any knowledge of the signal, one 
may have to resort to classical techniques such as 
energy detection \cite{poor}. An energy detector may need to collect data over a long 
period of time to detect the primary users reliably. 
Moreover, controlling the false alarm rates in mobile applications is difficult because the statistics of the signals, noise and interference may be time-varying. 
Another significant drawback is that energy detection has no 
capability to distinguish among different types of transmissions 
or to dichotomize between primary and secondary users of the spectrum.

Cyclostationary processes are random processes for which the statistical 
properties such as the mean and 
autocorrelation change periodically as functions of time \cite{gardner86}. 
Many of the signals used in 
wireless communication and radar systems possess this property. Cyclostationarity may be 
caused by modulation or coding \cite{gardner86}, or it may be also intentionally produced in order to 
aid channel estimation or synchronization \cite{tsatsanis}. 
Cyclostationarity property has been widely used in intercept 
receivers \cite{gardner86,koivisto,lunden}, 
direction of arrival or time-delay estimation, blind equalization and 
channel estimation \cite{tong} as well as in precoder design 
in multicarrier communications \cite{tsatsanis}. In order to exploit cyclic statistics, 
the signal must
be oversampled with respect to the symbol rate, or multiple receivers must be 
used to observe the signal.
The use of cyclostationary statistics is appealing in many ways: noise is rarely cyclostationary and 
second-order cyclostationary statistics retain also the phase information. 
Hence, procedures based on cyclostationarity tend to have particularly good performance at the low SNR regime. Moreover, cyclostationarity allows 
for distinguishing among different transmission types and users if 
their signals have distinct cyclic frequencies. A comprehensive list of references on cyclostationarity along with a survey of the literature is presented in~\cite{gardner06}.

The presence of cyclostationary signals may be determined by using hypothesis testing. 
Many existing tests, such as \cite{dandawate94}, 
are able to detect the presence of cyclostationarity at only one 
cyclic frequency at a time,
and they partly ignore the rich information present in the signals. For example, 
a communication signal may have cyclic frequencies related to the carrier frequency, 
the symbol rate and its harmonics, the chip rate, guard period, the scrambling code period, and the channel coding  scheme. 
In this paper we propose a method for detecting multiple 
cyclic frequencies simultaneously. It extends the method of \cite{dandawate94} 
to take into account the rich information present at different cyclic frequencies. 
This provides improved detector performance over techniques relying only on 
single cyclic frequency and facilitates dichotomizing among the 
primary and secondary user signals and different waveforms used.

In cognitive radio systems, there are typically multiple geographically distributed 
secondary users (SUs) that need to detect if the primary user is transmitting. 
The distributed sensors may work collaboratively to decide between two hypotheses: 
is the primary user active, or is the spectrum unused and available for the secondary 
users?  Decentralized processing has a number of advantages for such situations. Obviously, it allows for a larger coverage area. Furthermore, there are gains similar to diversity gains in wireless communications so that
the detection becomes less sensitive to demanding propagation conditions 
such as shadowing by large obstacles, 
large numbers of scatterers, differences in attenuation, or fast fading 
caused by mobility. Moreover, 
distributed sensory systems may require less communication bandwidth, 
consume less power, 
be more reliable and cost less as well. In this paper, we propose a simple 
decentralized decision making approach based on sharing and combining quantized 
local decision statistics. This approach may be used in both decision 
making with or without a fusion center.

This paper is organized as follows. In Section II, there is a short review of cyclostationary statistics. A novel detector for multiple cyclic frequencies is derived in Section III.  Section IV addresses the problem of collaborative detection of primary user. Simulation results demonstrating the detector's reliability in the low SNR regime 
as well as the gains obtained via collaborative operation are presented in Section V.
Finally, conclusions are drawn in Section VI.

\section{Cyclostationarity: a recap}

\label{sec:cyclo_def}

In this section, we provide a brief overview of cyclostationarity 
in order to make the derivation of the detector in Section III clearer. A continuous-time
random process  $x(t)$ is wide sense second-order cyclostationary if there exists a $T_0>0$ such 
that \cite{gardner86}:
\begin{equation}
\mu_{x}(t)   =  \mu_{x}(t+T_0)\ \forall t \\
\end{equation}
and
\begin{equation}
R_{x}(t_{1},t_{2})  =  R_{x}(t_{1}+T_0,t_{2}+T_0)\ \forall t_{1},t_{2}.
\label{eq:csdef}
\end{equation}
$T_0$ is called the period of the cyclostationary process. 

Due to the periodicity of the autocorrelation $R_{x}(t_{1},t_{2})$, it has a Fourier-series representation.
By denoting $t_{1}=t+\tau/2$ and $t_{2}=t-\tau/2$, we obtain the following expression for the 
Fourier-series \cite{gardner86}:
\begin{equation}
R_{x}(t+\frac{\tau}{2},t-\frac{\tau}{2})=\sum_{\alpha}R_{x}^{\alpha}(\tau)e^{j2\pi\alpha t},
\label{eq:caf_series}
\end{equation}
where the Fourier coefficients are
\begin{equation}
R_{x}^{\alpha}(\tau)=\frac{1}{T_0}\int_{-\infty}^{\infty}R_{x}(t+\frac{\tau}{2},t-\frac{\tau}{2})e^{-j2\pi\alpha t}dt
\label{eq:caf}
\end{equation}
and $\alpha$ is called the cyclic frequency. 
The function $R_{x}^{\alpha}(\tau)$ is called the cyclic autocorrelation function. 
If the process has zero mean, then this is also the cyclic autocovariance function.

When the autocorrelation function has exactly one period $T_0$
we have the following set of cyclic frequencies
\[
{\mathcal A}=\left\{\alpha=k/T,k\geq1 \right\},
\]
where $R_{x}^{\alpha}(\tau)$ is the cyclic autocorrelation function
and $\mathcal{A}$ are the set of cyclic frequencies. The
cyclic frequencies are harmonics of the fundamental frequency. 
If the autocorrelation function has several periods $T_{0},T_{1},\ldots$, 
we may express $R_{x}^{\alpha}(\tau)$ at the limit \cite{gardner86}
\begin{equation}
R_{x}^{\alpha}(\tau)=\lim_{T\rightarrow\infty}\frac{1}{T}\int_{-T/2}^{T/2}x(t+\frac{\tau}{2})x^{\ast}(t-\frac{\tau}{2})e^{j2\pi\alpha t}dt.
\end{equation}
The process $x(t)$ is almost cyclostationary in the wide sense and the 
set of cyclic frequencies ${\mathcal A}$ is comprised of a countable number of 
frequencies that do not need to be harmonics of the fundamental frequency.
In general, the process is said to be cyclostationary if there exists an $\alpha\neq 0$ such that $R_{x}^{\alpha}(\tau)\neq 0$ for some value of $\tau$. Typically cyclic frequencies are assumed to be known or 
may be estimated reliably.

\section{Detection using multiple cyclic frequencies}

Statistical tests for the presence of a single cyclic frequency have been proposed, 
for example, in~\cite{dandawate94}. 
The tests in~\cite{dandawate94} have asymptotically 
constant false alarm rate (CFAR) for testing presence of cyclostationarity at a
given cyclic frequency.  However, the tests do not retain the CFAR property over 
a set of tested frequencies.
 
Typical communication signals exhibit cyclostationarity at multiple cyclic 
frequencies instead of just a single cyclic frequency.
That is, for example a signal that is cyclostationary at the symbol frequency is 
typically cyclostationary at all integer multiples of the symbol frequency as well. 
There also may be cyclic frequencies related to the coding and  guard periods, 
or adaptive modulation and coding may be used. In such cases the cyclic frequencies present 
may vary depending on channel quality and the waveform used.
If one is testing for the presence of many different signals at a given frequency
band, or in case the cyclic frequencies are not known, 
it would be desirable to retain the CFAR property over the whole set of tested cyclic
frequencies. This would be especially desirable in a cognitive radio 
application where the interest is in finding unoccupied
frequency bands. Otherwise the frequency band may unnecessarily be 
classified as occupied for most of the time.
 
In the following we extend the test based on second-order cyclic statistics 
of~\cite{dandawate94} to multiple cyclic frequencies. 
To do so we first define all the terms used in the test statistics. 

Let $(*)$ denote an optional complex conjugation. The notation allows convenient handling of both cyclic autocorrelation and conjugate cyclic autocorrelation with only one equation. An estimate of 
the (conjugate) cyclic autocorrelation ${\hat R}_{xx^{(*)}}(\alpha,\tau)$ may be obtained using $M$ observations as
\begin{align}
\label{eq:cyclic_autocorr_estimator}
{\hat R}_{xx^{(*)}}(\alpha,\tau) &= 
\frac{1}{M} \sum_{t=1}^{M} x(t)x^{(*)}(t+\tau) e^{-j 2 \pi \alpha t}\\
&= R_{xx^{(*)}}(\alpha,\tau) + \varepsilon(\alpha,\tau),
\end{align}
where the latter term is the estimation error. This estimator is consistent, (see ~\cite{dandawate94}) 
so that the error goes to zero as $M \rightarrow \infty$.

Now we need to construct a test for a number of lags $\tau_1,\ldots,\tau_{N}$ as well as a set of cyclic frequencies of interest. Let $\mathcal{A}$ denote the set of cyclic frequencies of interest, and
\begin{equation}
\begin{aligned}
\hat{\bm{r}}_{xx^{(*)}}(\alpha) = \bigg[& \mathrm{Re}\{\hat{R}_{xx^{(*)}}(\alpha,\tau_{1})\},\ldots,\mathrm{Re}\{\hat{R}_{xx^{(*)}}(\alpha,\tau_{N})\}, \\
&\mathrm{Im}\{\hat{R}_{xx^{(*)}}(\alpha,\tau_{1})\},\ldots,\mathrm{Im}\{\hat{R}_{xx^{(*)}}(\alpha,\tau_{N})\} \bigg]
\end{aligned}
\end{equation}
denote a $1 \times 2N$ vector containing the real and
imaginary parts of the estimated cyclic autocorrelations at the cyclic frequency of interest stacked in a single vector.

The $2N \times 2N$ covariance matrix of $\bm{r}_{xx^{(*)}}$ can be computed as~\cite{dandawate94} 
\begin{equation}
\bm{\Sigma}_{xx^{(*)}}(\alpha) = \begin{bmatrix} \mathrm{Re}\left\{ \frac{\bm{Q}+\bm{Q}^{*}}{2}\right\} &  \mathrm{Im}\left\{ \frac{\bm{Q}-\bm{Q}^{*}}{2}\right\} \\ 
\mathrm{Im}\left\{ \frac{\bm{Q}+\bm{Q}^{*}}{2}\right\} &  \mathrm{Re}\left\{ \frac{\bm{Q}^{*}-\bm{Q}}{2}\right\} \end{bmatrix}
\end{equation}
where the $(m,n)$th entries of the two covariance matrices $\bm{Q}$ and $\bm{Q}^{*}$ are given by
\[
\begin{aligned}
\bm{Q}(m,n) &= S_{f_{\tau_{m}}f_{\tau_{n}}}(2\alpha,\alpha)
\end{aligned}
\]
and
\begin{equation}
\begin{aligned}
\bm{Q}^{*}(m,n) &= S^{*}_{f_{\tau_{m}}f_{\tau_{n}}}(0,-\alpha).
\end{aligned}
\end{equation}
Here, $S_{f_{\tau_{m}}f_{\tau_{n}}}(\alpha,\omega)$ and $S^{*}_{f_{\tau_{m}}f_{\tau_{n}}}(\alpha,\omega)$ denote the unconjugated and conjugated cyclic spectra of $f(t,\tau) = x(t)x^{(*)}(t+\tau)$, respectively. 
These spectra can be estimated using frequency smoothed cyclic periodograms as
\begin{align}
\hat{S}_{f_{\tau_{m}}f_{\tau_{n}}}(2\alpha,\alpha) &= 
\frac{1}{ML} \sum_{s=-(L-1)/2}^{(L-1)/2} W(s) \nonumber \\
&\mbox{} \quad \cdot F_{\tau_{n}}(\alpha-\frac{2\pi s}{M})F_{\tau_{m}}(\alpha+\frac{2\pi s}{M}) \\
\hat{S}^{*}_{f_{\tau_{m}}f_{\tau_{n}}}(0,-\alpha) &= 
\frac{1}{ML} \sum_{s=-(L-1)/2}^{(L-1)/2}W(s) \nonumber \\
&\mbox{} \quad \cdot F^{*}_{\tau_{n}}(\alpha+\frac{2\pi s}{M})F_{\tau_{m}}(\alpha+\frac{2\pi s}{M})
\end{align}
where $F_{\tau}(\omega) = \sum_{t=1}^{M}x(t)x^{(*)}(t+\tau)e^{-j\omega t}$ and $W$ is a 
normalized spectral window of odd length $L$.

Now the hypothesis testing problem for testing if $\alpha$ is a cyclic frequency can be formulated as~\cite{dandawate94}
\begin{equation}
\label{eq:hypothesis_single}
\begin{aligned}
H_{0}: &\mbox{} \;\forall \{\tau_{n}\}_{n=1}^{N} \Longrightarrow \hat{\bm{r}}_{xx^{(*)}}(\alpha) = \bm{\epsilon}_{xx^{(*)}}(\alpha) \\
H_{1}: &\mbox{}\; \mathrm{for}\;\mathrm{some}\; \{\tau_{n}\}_{n=1}^{N} \\ &\Longrightarrow \hat{\bm{r}}_{xx^{(*)}}(\alpha) = \bm{r}_{xx^{(*)}}(\alpha) + \bm{\epsilon}_{xx^{(*)}}(\alpha). 
\end{aligned}
\end{equation}
Here $\bm{\epsilon}_{xx^{(*)}}$ is the estimation error which is asymptotically normal distributed, i.e., $\lim_{M\rightarrow \infty} \sqrt{M}\bm{\epsilon}_{xx^{(*)}} \overset{D}{=} N(\bm{0},\bm{\Sigma}_{xx^{(*)}})$~\cite{dandawate94}. 
Hence, using the asymptotic normality of $\hat{\bm{r}}_{xx^{(*)}}$ the generalized likelihood ratio (GLR) is given by
\begin{equation}
\begin{aligned}
\Lambda &= \frac{\exp(-\frac{1}{2}M\hat{\bm{r}}_{xx^{(*)}}\hat{\bm{\Sigma}}_{xx^{(*)}}^{-1}\hat{\bm{r}}_{xx^{(*)}}^{T})}{\exp(-\frac{1}{2}M(\hat{\bm{r}}_{xx^{(*)}}-\hat{\bm{r}}_{xx^{(*)}})\hat{\bm{\Sigma}}_{xx^{(*)}}^{-1}(\hat{\bm{r}}_{xx^{(*)}}-\hat{\bm{r}}_{xx^{(*)}})^{T})} \\
&=\exp(-\frac{1}{2}M\hat{\bm{r}}_{xx^{(*)}}\hat{\bm{\Sigma}}_{xx^{(*)}}^{-1}\hat{\bm{r}}_{xx^{(*)}}^{T}). \\
\end{aligned}
\end{equation} 

Finally, by taking the logarithm and multiplying the result by 2, 
we arrive at the test statistic in~\cite{dandawate94}
\begin{equation}
\label{eq:cyclic_test_statistic}
\mathcal{T}_{xx^{(*)}}(\alpha) = -2\ln \Lambda = M\hat{\bm{r}}_{xx^{(*)}}\hat{\bm{\Sigma}}_{xx^{(*)}}^{-1}\hat{\bm{r}}_{xx^{(*)}}^{T}.
\end{equation}
Under the null hypothesis $\mathcal{T}_{xx^{(*)}}(\alpha)$ is asymptotically $\chi_{2N}^{2}$ 
distributed.

Now in order to extend the test for the presence of second-order cyclostationarity at any of the 
cyclic frequencies of interest $\alpha \in \mathcal{A}$ simultaneously, we formulate the hypothesis 
testing as follows
\begin{equation}
\label{eq:hypothesis_multi}
\begin{aligned}
H_{0}: &\mbox{}\;\forall \alpha \in \mathcal{A}\;\mathrm{and} \;\forall \{\tau_{n}\}_{n=1}^{N} \Longrightarrow \hat{\bm{r}}_{xx^{(*)}}(\alpha) = \bm{\epsilon}_{xx^{(*)}}(\alpha) \\
H_{1}: &\; \mathrm{for}\;\mathrm{some}\;\alpha \in \mathcal{A}\; \mathrm{and}\;\mathrm{for}\;\mathrm{some}\; \{\tau_{n}\}_{n=1}^{N}\\
& \Longrightarrow \hat{\bm{r}}_{xx^{(*)}}(\alpha) = \bm{r}_{xx^{(*)}}(\alpha) + \bm{\epsilon}_{xx^{(*)}}(\alpha). 
\end{aligned}
\end{equation}

For this detection problem, we propose the following two test statistics:
\begin{align}
\mathcal{D}_{m} &= \max_{\alpha \in \mathcal{A}} \mathcal{T}_{xx^{(*)}}(\alpha)
= \max_{\alpha \in \mathcal{A}} M\hat{\bm{r}}_{xx^{(*)}}(\alpha)
\hat{\bm{\Sigma}}_{xx^{(*)}}^{-1}(\alpha)\hat{\bm{r}}_{xx^{(*)}}^{T}(\alpha) \\
\mathcal{D}_{s} &= \sum_{\alpha \in \mathcal{A}} \mathcal{T}_{xx^{(*)}}(\alpha)
= \sum_{\alpha \in \mathcal{A}} M\hat{\bm{r}}_{xx^{(*)}}(\alpha)
\hat{\bm{\Sigma}}_{xx^{(*)}}^{-1}(\alpha)\hat{\bm{r}}_{xx^{(*)}}^{T}(\alpha).
\end{align}

The first test statistic calculates the maximum of the cyclostationary GLRT statistic~\eqref{eq:cyclic_test_statistic} over the cyclic frequencies of interest $\mathcal{A}$ while the second calculates the sum. Assuming independence of cyclic autocorrelation estimates for different cyclic frequencies the test statistic $\mathcal{D}_{s}$ is the GLRT statistic. Depending on the signal and the set of tested cyclic frequencies the test statistics may have different performances. This requires further research.

The asymptotic distribution of $\mathcal{D}_{s}$ is under the null hypothesis $\chi_{2NN_{\alpha}}^{2}$ where $N_{\alpha}$ is the number of cyclic frequencies in set $\mathcal{A}$. This is due to the fact that the sum of independent 
chi-square random variables is also a chi-square random variable whose degrees of freedom is 
the sum of the degrees of freedom of the independent random variables.

In the following we derive the asymptotic distribution of the test statistic $\mathcal{D}_{m}$ under the null hypothesis. As stated above, under the null hypothesis $\mathcal{T}_{xx^{(*)}}(\alpha)$ is asymptotically $\chi_{2N}^{2}$ distributed.
The cumulative distribution function of the chi-square distribution with $2N$ degrees of freedom is given by
\begin{equation}
F(x,2N) = \frac{\gamma (N,x/2)}{\Gamma (N)}
\end{equation}
where $\gamma(k,x)$ is the lower incomplete gamma function and $\Gamma(k)$ is the ordinary gamma function. For a positive
integer $k$ the following identities hold:
\begin{align}
\Gamma(k) &= (k-1)! \\
\gamma(k,x) &= \Gamma(k) - (k-1)! \; e^{-x}\sum_{n=0}^{k-1}\frac{x^{n}}{n!}. 
\end{align}

Hence, the cumulative distribution function of the chi-square distribution with $2N$
degrees of freedom is given by
\begin{equation}
F(x,2N) = 1-e^{-x/2}\sum_{n=0}^{N-1}\frac{(x/2)^{n}}{n!}.
\end{equation}
 
The cumulative distribution function of the maximum of $d$ 
independent and identically distributed random variables is the cumulative
distribution function of the individual random variables raised to the power $d$. 
Thus, the cumulative distribution function of the test statistic $\mathcal{D}_{m}$
is given by
\begin{equation}
F_{\mathcal{D}_{m}}(x,2N,d) = \left(1-e^{-x/2}\sum_{n=0}^{N-1}\frac{(x/2)^{n}}{n!}\right)^{d}.
\end{equation}  

The corresponding probability density function is obtained by differentiating 
the cumulative distribution function, i.e.,
\begin{equation}
\begin{aligned}
f_{\mathcal{D}_{m}}(x,2N,d) &= \frac{d}{2}\left(1-e^{-x/2}\sum_{n=0}^{N-1}\frac{(x/2)^{n}}{n!}\right)^{d-1} \\
&\mbox{} \quad \cdot e^{-x/2}\left(\sum_{n=0}^{N-1}\frac{(x/2)^{n}}{n!}-\sum_{n=1}^{N-1}\frac{(x/2)^{n-1}}{(n-1)!}\right).
\end{aligned}
\end{equation}
 
Consequently, the null hypothesis is rejected if $F_{\mathcal{D}_{m}}(\mathcal{D}_{m},2N,N_{\alpha}) > 1-p$ where $p$ is the false alarm rate and $N_{\alpha}$ is the number of tested cyclic frequencies.

\section{Cooperative detection}

User cooperation may be used to improve the performance and coverage in a cognitive radio network. 
The users may collaborate in finding unused spectrum and new opportunities. Many of the 
collaborative detection techniques stem from distributed detection theory; see \cite{viswanathan,blum}.
In cognitive radio systems, there are typically multiple geographically 
distributed secondary users that need to detect 
whether the primary user is active. All the secondary users may sense the 
entire band of interest, or monitor just a partial band to reduce power consumption. 
In the latter case each SU senses a certain part of the spectrum, 
and then shares the acquired information 
with other users or a fusion center. 

The cooperation may then be coordinated by a fusion center (FC), 
or it may take place in an ad-hoc manner 
without a dedicated fusion center. Here we assume that a fusion center collects information from 
all $K$ secondary users and makes a decision about whether the spectrum is available or not. 
We assume that each secondary user sends a quantized version of its local decision statistics 
(such as the likelihood ratio) to the FC. In the case of very coarse quantization,  
binary local decision may be sent. 
To derive a test for the FC, 
we assume that the sensors are 
independent conditioned on whether the hypothesis $H_0$ or $H_1$ is true. 
Then the optimal fusion rule is the likelihood ratio test over 
the received local likelihood ratios $l_i$:
\begin{equation}
\mathcal{T}_{K} = \prod_{i=1}^{K} l_i. 
\end{equation}
In case the secondary users send binary decisions, the sum of ones may calculated and compared to a threshold. 
Here, we consider the simplest way of making the decision using generalized likelihood ratios. 
Instead of using the product of the generalized likelihood ratios, 
we can employ the sum of generalized log-likelihood ratios. We propose the following test statistic for the hypothesis testing problem~\eqref{eq:hypothesis_single} 
\begin{equation}
\mathcal{T}_{K}' = \sum_{i=1}^{K} {\mathcal{T}^{(i)}_{xx^{(*)}}(\alpha)},
\end{equation}
and the following two for the hypothesis testing problem~\eqref{eq:hypothesis_multi}
\begin{align}
\mathcal{D}_{m,K} &= \max_{\alpha \in \mathcal{A}} \sum_{i=1}^{K} {\mathcal{T}^{(i)}_{xx^{(*)}}(\alpha)} \\
\mathcal{D}_{s,K} &= \sum_{\alpha \in \mathcal{A}} \sum_{i=1}^{K} {\mathcal{T}^{(i)}_{xx^{(*)}}(\alpha)}
\end{align}
where ${\mathcal{T}^{(i)}_{xx^{(*)}}(\alpha)}$ is the cyclostationarity 
based test statistic~\eqref{eq:cyclic_test_statistic} from $i^{th}$ secondary user. Due to the use of generalized likelihood ratios, no optimality properties can be claimed. 
The GLRT test does, however, perform highly reliably in many applications.

Under the conditional independence assumption the asymptotic distributions of the test statistic 
$\mathcal{T}_{K}'$ and $\mathcal{D}_{s,K}$ are under the null hypothesis $\chi_{2NK}^{2}$ and $\chi_{2NN_{\alpha}K}^{2}$, respectively. This is again due to the fact that the sum of independent 
chi-square random variables is also a chi-square random variable whose degrees of freedom is 
the sum of the degrees of freedom of the independent random variables. The cumulative distribution function of $\mathcal{D}_{m,K}$ is under the null hypothesis $F_{\mathcal{D}_{m}}(\mathcal{D}_{m,K},2NK,N_{\alpha})$ where $N_{\alpha}$ is again the number of tested cyclic frequencies. The testing is done similarly as in one secondary user case.

Different techniques for reducing the amount of transmitted data, taking into account the 
relevance of the information provided by secondary users as well as how to  
deal with communication rate constraints will be addressed in a forthcoming paper.

\section{Simulation examples}

In this section the performance of the proposed detectors is considered. 
The test signal is an orthogonal frequency division multiplex (OFDM) signal. 
The baseband equivalent of a cyclic prefix OFDM signal may be expressed as 
\begin{equation}
x(t) = \sum_{n=0}^{N_{c}-1}\sum_{l=-\infty}^{\infty}c_{n,l}g(t-lT_{s})e^{j(2\pi/N)n(t-lT_{s})}
\end{equation}
where $N_{c}$ is the number of subcarriers, $T_{s}$ is the symbol length, $g(t)$ denotes the rectangular pulse of length $T_{s}$, and the $c_{n,l}$'s denote the data symbols. 
The symbol length is the sum of the length of the useful symbol data $T_{d}$ and the length of the 
cyclic prefix $T_{cp}$, i.e., $T_{s} = T_{d} + T_{cp}$.  

The above OFDM signal exhibits cyclostationarity (i.e., complex conjugation is used in~\eqref{eq:cyclic_autocorr_estimator} and the following equations) with 
cyclic frequencies of $\alpha = k/T_{s}$, $k = 0, \pm 1, \pm 2, \ldots$ and potentially other frequencies depending on the coding scheme. The cyclic autocorrelation surfaces for $\alpha = k/T_{s}$ peak at $\tau = \pm T_{d}$~\cite{oner04}. 

In the following the performance of cyclic detectors based on one and two cyclic frequencies is compared as a function of signal-to-noise ratio (SNR) in an additive white Gaussian noise (AWGN) channel. 
The SNR is defined as $\mathrm{SNR} = 10\log_{10}\frac{\sigma_{x}^{2}}{\sigma_{n}^{2}}$ where $\sigma_{x}^{2}$ and $\sigma_{n}^{2}$ are the variances of the signal and the noise, respectively. The cyclic frequencies employed by the detectors are $1/T_{s}$ and $2/T_{s}$. The detector based on one cyclic frequency uses the first frequency and the detectors based on two cyclic frequencies use both frequencies. Each detector uses 
two time lags $\pm T_{d}$.  

The cyclic spectrum estimates were calculated using a length-2049 Kaiser window with $\beta$ parameter of 10. A Fast-Fourier transform (FFT) was employed for faster computation. 
The FFT size was 10000 giving a cyclic frequency resolution of 0.0001.

The OFDM signal has 32 subcarriers and the length of the cyclic prefix is 1/4 of the useful symbol data. The subcarrier modulation employed is 16-QAM. The signal length is 100 OFDM symbols. 
 
\begin{figure}[tbp]
	\centering
		\includegraphics[width=0.75\columnwidth]{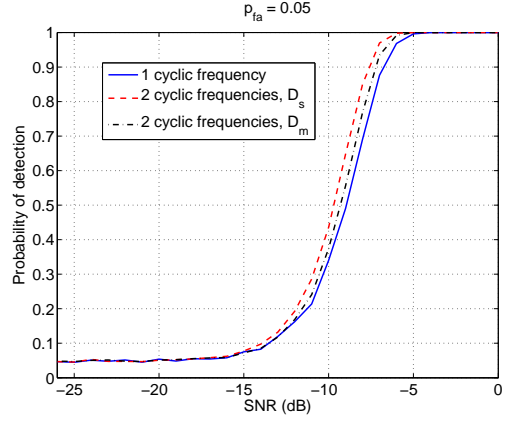}
	\caption{Probability of detection vs. SNR. The multicycle detectors achieve better performance than the single cycle detector in the low SNR regime. The sum detector of the test statistic $\mathcal{D}_{s}$ has the best performance.}
	\label{fig:ofdm_pd_vs_SNR}	
\end{figure}

\begin{figure}[tbp]
	\centering
		\includegraphics[width=0.75\columnwidth]{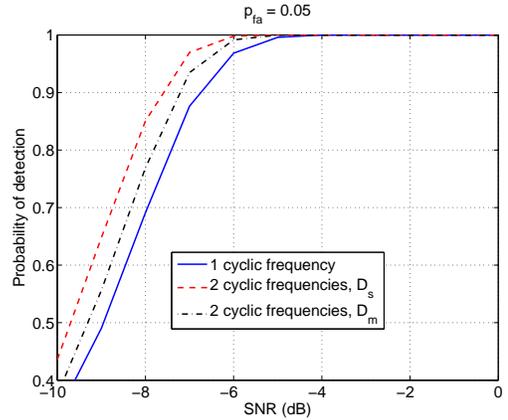}
	\caption{Probability of detection vs. SNR. Zoom of the important region. The multicycle detectors achieve better performance than the single cycle detector in the low SNR regime. The sum detector of the test statistic $\mathcal{D}_{s}$ has the best performance.}	
	\label{fig:ofdm_pd_vs_SNR_zoom}	
\end{figure}

Fig.~\ref{fig:ofdm_pd_vs_SNR} depicts the performance of the detectors as a function of the SNR for a constant false alarm rate of 0.05. Fig.~\ref{fig:ofdm_pd_vs_SNR_zoom} shows a zoom of the important area illustrating the differences in performance more clearly. All the curves are averages over 10000 experiments. It can be seen that the detectors based on multiple cyclic frequencies outperform the detector based on single cyclic frequency in the low SNR regime. Furthermore, the multicycle detector calculating the sum over the cyclic statistics of different frequencies has the best performance.
 
\begin{figure}[tbp]
	\centering
		\includegraphics[width=0.75\columnwidth]{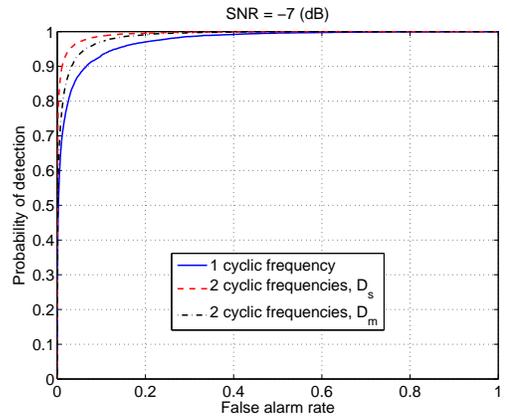}
	\caption{Probability of detection vs. false alarm rate. The detectors based on multiple cyclic frequencies achieve better performance than the detector based on a single cyclic frequency.}
	\label{fig:ofdm_pd_vs_pfa_1}	
\end{figure}

Fig.~\ref{fig:ofdm_pd_vs_pfa_1} plots the probability of detection vs. false alarm rate for SNR of -7 dB. The figure show that the detectors have desirable receiver operating characteristics. That is, the probability of detection increases as the false alarm rate parameter is increased.
 
Next the performance gain from cooperative detection of several secondary users is analyzed. The signal is the same as above. The cooperative detection is based on the data of 5 secondary users. Each secondary user receives the same data with different noise. SNR is the same for each secondary user. 

\begin{figure}[tbp]
	\centering
		\includegraphics[width=0.75\columnwidth]{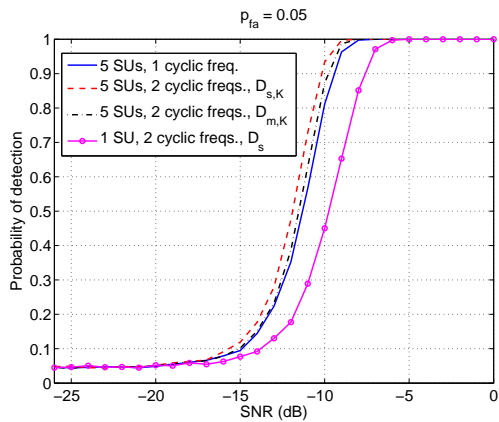}
	\caption{Probability of detection vs. SNR. Cooperation of 5 secondary users provides performance gain of 3 dB. Using multiple cyclic frequencies further improves the detection performance. The sum detector of the test statistic $\mathcal{D}_{s,K}$ has the best performance.}
	\label{fig:ofdm_5su_pd_vs_SNR}	
\end{figure}

\begin{figure}[tbp]
	\centering
		\includegraphics[width=0.75\columnwidth]{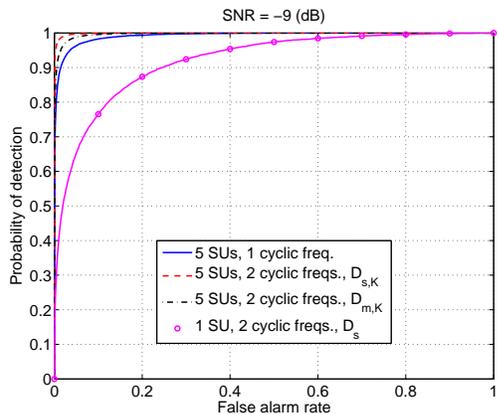}
	\caption{Probability of detection vs. false alarm rate. Cooperation among secondary users combined with the use of multicycle sum test statistic $\mathcal{D}_{s,K}$ provides the best performance.}
	\label{fig:ofdm_5su_pd_vs_pfa}	
\end{figure}

Fig.~\ref{fig:ofdm_5su_pd_vs_SNR} depicts the performance for 5 secondary users compared to the single secondary user case. Performance gain of roughly 3 dB is obtained from the cooperation of 5 secondary users. Using two cyclic frequencies provides similar performance improvement as in single secondary user case. Fig.~\ref{fig:ofdm_5su_pd_vs_pfa} shows the probability of detection vs. false alarm rate for SNR of -9 dB.

\begin{figure}[tbp]
	\centering
		\includegraphics[width=0.8\columnwidth]{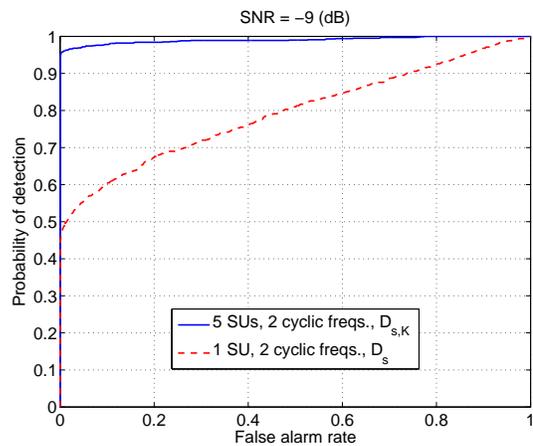}
	\caption{Probability of detection vs. false alarm rate. In order to simulate shadowing the SNR of each user was independently selected randomly from a normal distribution with a mean of -9 dB and standard deviation of 10 dB. Cooperation among secondary users reduces sensitivity to shadowing effects.}
	\label{fig:ofdm_shadowing}	
\end{figure}

In the following simplistic example, we illustrate the gains that may be achieved 
via collaborative detection in the face of shadowing effects. 
In order to simulate shadowing, the SNR of each user was independently selected randomly from a 
normal distribution with a mean of -9 dB and standard deviation of 10 dB. 
That is, the logarithm of the received power level is normally distributed. 
Fig.~\ref{fig:ofdm_shadowing} depicts the performance of the multicycle detectors for the 
simple shadowing scenario. Comparison to Fig.~\ref{fig:ofdm_5su_pd_vs_pfa} reveals that cooperation among secondary users reduces sensitivity to shadowing effects significantly.
 
\section{Conclusion}

In this paper, a generalized likelihood ratio test for detecting primary transmissions 
with multiple cyclic frequencies has been proposed, and the 
asymptotic distribution of the test statistic has been derived.
In this test, impairments such as shadowing and fading are mitigated by
combining the quantized local likelihood ratios from a number of secondary users under 
a conditional independence assumption. Simulation examples 
demonstrating the improved reliability in the detector performance in the low SNR regime as 
well as significant gains obtained via collaborative decision making have also been presented.

\end{document}